# Bot or Not? Deciphering Time Maps for Tweet Interarrivals

Nicole M. Radziwill & Morgan C. Benton

**Abstract**: This exploratory study used the R Statistical Software to perform Monte Carlo simulation of time maps, which characterize events based on the elapsed time since the last event and the time that will transpire until the next event, and compare them to time maps from real Twitter users. Time maps are used to explore differences in the interarrival patterns of Tweets between human users, humans who use scheduling services like TweetDeck and HootSuite, and non-human ("bot") users. The results indicate that there are differences between the tweet interarrival patterns across these categories of users, and that time maps could potentially be used to automate the detection of bot accounts on Twitter. This could enhance social media intelligence capabilities, help bot developers build more "human-like" Twitter bots to avoid detection, or both.

**Introduction and Background**

Many phenomena can be represented as ordered sequences of well-defined (or "discrete" events). These include arrivals, departures, and occurrences, and can represent occasions as diverse as notable weather events (e.g. hail, tornadoes), use of a machine, delivery of a service, phone calls, text messages, posts to discussion boards, and 140-character or less messages posted to the Twitter service at http://twitter.com, called "tweets.".

This paper uses the R Statistical Software (R Core Team, 2015) to perform Monte Carlo simulation to explore the patterns that arise in time maps, a new data visualization technique (Watson, 2015) that can be used to explore patterns in the interarrival times between events across multiple timescales, including tweets. Five distributions of interarrival times are considered: exponential, uniform, Gaussian, and a mixture distribution that resembles a "hierarchically bundled" process. These results are compared to interarrival patterns observed on real Twitter accounts to demonstrate differences between human users, human users who employ scheduling services like TweetDeck and HootSuite, and non-human ("bot") users.

Real-time information is critical if a decision-maker needs to assess people and situations to produce actionable intelligence, particularly in cases of national security. (Ivan, 2015) One source of plentiful real-time information is Twitter, a social media website with over 320 million users as of December 2015. (Twitter, 2015) On Twitter, users construct and broadcast short (<140 character) messages called "tweets" which can be accessed and mined in real-time. Research to date revolving around Twitter data streams has focused on event detection (using tweet streams to detect whether larger-scale events like earthquakes or protests have occurred) or event prediction (extracting the likelihood of events forming or developing based on user expressions or sentiments). For example, Hidden Markov Models have been used to explore the feasibility of reconstructing event summaries from tweet streams, a feature that may be evident now in Twitter's "Moments" tab. (Chakrabarti & Punera, 2011)

The state of research with respect to *using* the tweet stream to create intelligence that is useful for decision-making is, however, still immature. One step towards improving the quality of information from the tweet stream is being able to detect what *type* of account a tweet is coming from: a human, a human whose tweets are scheduled rather than spontaneous, or an autonomous bot whose tweets may not fit either pattern. This distinction is important because humans are excellent at mining subtle social and emotional cues to distinguish significant, high-impact news and events from ordinary information, whereas machines are less effective. (Petrović, Osborne, & Lavrenko, 2010)

Tweets are discrete events, and tweet interarrivals (the times between successive Tweets) have successfully been modeled as homogeneous Poisson processes (Gonzalez, Muñoz, & Hernández, 2014) and inhomogeneous log-Gaussian Cox processes. (Lukasic et al., 2015) The bursty nature of human-initiated tweets has also been associated with non-Poisson interarrivals, and attributed to "decision-based queuing processes, where individuals tend to act in response to some perceived priority." (De Domenico, 2013) However, to the authors' knowledge, no studies to date seek to distinguish the account type by exploring interarrival patterns. The primary contribution of the simulation below is to take the first step towards determining the feasibility of account discrimination on Twitter for this purpose using time maps.

**Methods**

This study uses Monte Carlo simulation to explore the structure of time maps for various distributions of Tweet interarrival times, and compares those idealized maps to real maps from Twitter accounts. This information is used to determine whether a qualitative difference is apparent between human users, humans who schedule their tweets, and bots.

Time maps, similar to phase space diagrams, present all observed combinations of timing sequences by plotting points whose coordinates are the *time prior to* and *time subsequent to* an observed tweet. Consequently, this analysis can only be done in near real-time, since knowledge of a future state is required from each event. Each point represents one event (in this case, a tweet) that is characterized by the time that has elapsed since the previous event (its x-coordinate) and the time that will elapse before the next event (its y-coordinate), as shown in Figure 1. The map is log-scaled to enhance the visibility of all data points.

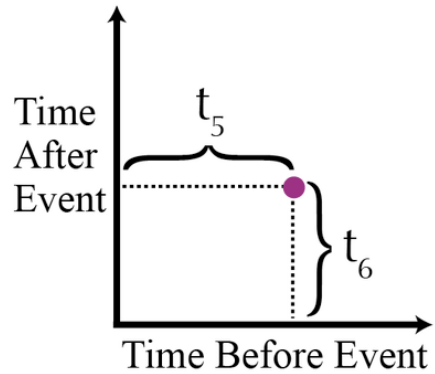

**Figure 1:** Points on time maps contain information about prior and subsequent events (from Watson, 2015)

The time maps emphasize the *relationships between events* rather than the events themselves. For example, if all observations in a stream of arrivals are spaced evenly so that the amount of time *before* the event is the same as the amount of time *after* the event, the time map generated from event clock times will be perfectly linear (Figure 2, left) while the time map generated from interarrival times will appear to contain only *one* point (Figure 2, right).

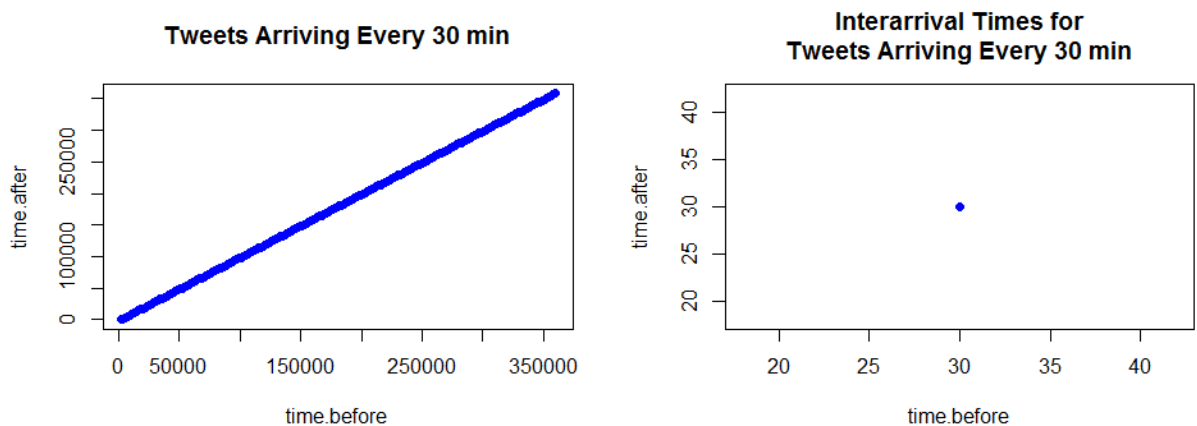

**Figure 2:** Example of a time map for clock times of arriving tweets (left) vs. interarrival times (right).

The steps involved in this study were:

1. **Select target users** in each of the following categories:
    a. Human users, with at least one:
        i. who tweets frequently but spontaneously.
        ii. who tweets in bursts after breaks of weeks to months.
        iii. who tweets infrequently but spontaneously.
    b. Human users with a strong social media presence, with at least one:
        i. who uses software to schedule tweets in advance.
        ii. who is popular in entertainment domains, rather than business.

  c. Known bot accounts with unknown tweeting strategies, with at least one:
    i. Autonomous bot
    ii. Reactive bot
    iii. Proactive bot
    iv. Social bot
    v. Adaptive bot
  d. 2016 U.S. Presidential Candidates, with at least one:
    i. who is a Republican.
    ii. who is a Democrat.

2. **Conduct Monte Carlo simulation** to determine the idealized characteristics of time maps for exponential, uniform, Gaussian, and lognormal distributions, plus a "hierarchically bundled" data generation process (DGP) at different sample sizes (n=10, 100, 1000, 10000)

3. **Obtain timing information** for up to 3200 tweets and retweets for each target user using the twitteR package in R.

4. **Construct time maps** for target users and conduct qualitative analysis to describe patterns.

Because the purpose of this study was to assess feasibility rather than to develop a robust, generalizable heuristic or classifier, no summary statistics will be provided for the time maps. If the results indicate feasibility, the next logical step will be to collect tweet interarrival times for a much larger sample of accounts within each of the user categories.

**Results and Discussion**

This section presents the following results: 1) a list of the Twitter accounts that were selected as "target users" for analysis, 2) time maps generated from simulated samples of tweets (n=10, 100, 1000, and 10000), 3) smoothed and unsmoothed log-transformed time maps for real samples from each of the target users, and 4) a table summarizing the qualitative assessment of patterns seen within the samples of tweet interarrivals.

*Selection of Target Users*

A total of 10 Twitter accounts were selected as target users (see Table 1). For group 1 (humans who do not schedule automated tweets), the authors selected themselves for convenience and interest. For group 2 (humans who do schedule automated tweets), the authors selected one friend and one acquaintance who are known to schedule tweets on a regular basis.

For group 3 (bots), the accounts were selected based on a list from Seward (2014), who did not perform a Turing test on each bot but did prequalify them based on levels of activity. From the

17 options, one Twitter bot was selected for each of the five dimensions of intelligent systems (autonomous, reactive, proactive, interactive/social, adaptive) and another bot was selected which did not appear to have characteristics of an intelligent system. (Franklin & Grasser, 1997)

For Group 4 (2016 U.S. Presidential Candidates), target users were selected based on the greatest number of Twitter followers. Although the intention was to cover both political parties, the selection was not related to the authors' political interests or any standings in the polls.

| Group | Twitter ID | Account Details |
| --- | --- | --- |
| **1: Humans who tweet spontaneously** | @nicoleradziwill | Author's account. She tweets about quality, data science, statistics, and random personal details. |
| | @morphatic | Author's account. He tweets mostly about programming and political issues. |
| | @[Confidential 1] & @[Confidential 2] | Real teenage girls. Included for comparison to @oliviataters, a "teenage girl bot" |
| | @larrysabato | Director, Univ. of Va. Center for Politics. Included because we suspect that he tweets spontaneously. |
| **2: Humans with a social media presence** | @marciamarcia | Authors' friend. She tweets about trends in business, creativity, innovation, and enterprise transformation. |
| | @valaafshar | Authors' acquaintance. He is the Chief Digital Evangelist at Salesforce.com and has nearly 100K followers. |
| | @dorieclark | Authors' acquaintance. She is an adjunct professor at Duke's Fuqua School of Business and author of *Stand Out*. |
| | @parishilton | Style icon, DJ, former socialite. Included because we suspect that she tweets spontaneously, and has 13.5M followers. |
| **3: Automated bots** | @dearassistant | **Adaptive**: Answers questions using Wolfram Alpha. |
| | @oliviataters | **Proactive**: Tweets using an algorithm that makes it sound like a teenage girl.. |
| | @a_quilt_bot | **Interactive**: Tweet an image at this account, and it will tweet back an image that has been modified using a "quilt" pattern. |
| | @reverseocr | **Autonomous**: Creates line drawings until Optical Character Recognition (OCR) software recognized |
| | @accidental575 | **Reactive**: Posts other tweets that are in the form of haiku. |
| | @netflix_bot | **Not intelligent**: Tweets information about new movies on Netflix. |
| **4: 2016 U.S. Presidential Candidates** | @berniesanders | **Bernie Sanders (D)** |

| | @hillaryclinton | **Hillary Clinton (D)** |
| --- | --- | --- |
| | @realdonaldtrump | **Donald Trump (R)** |
| | @realbencarson | **Dr. Ben Carson (R)** |
| | @carlyfiorina | **Carly Fiorina (R)** |
| | @tedcruz | **Ted Cruz (R)** |

**Table 1.** The Twitter accounts that were selected as "target users" for analysis.

*Time Maps With Simulated Interarrival Times*

Next, we generated time maps from simulated streams of data representing interarrival times randomly selected from data generating processes that *could* represent tweets: 1) a negative **exponential** distribution with a mean of 1 hr, 2) a **uniform** distribution ranging from 0 to 24 hrs, 3) a **Gaussian** distribution with a mean of 12 hrs and standard deviation of 3 hrs, and 4) a **mixture** distribution based on the "hierarchical bundling" of events that is observed in autocorrelated requests to Google datacenters, which exhibits burstiness. (Juan et al., 2014)

These distributions were selected not because they conform to a particular expectation for interarrival times, but because each data generating process produces interarrivals that *could* be reasonable for representing either human or bot behavior. The axes represent time, but only the patterns are physically meaningful in the simulations (and not the scales).

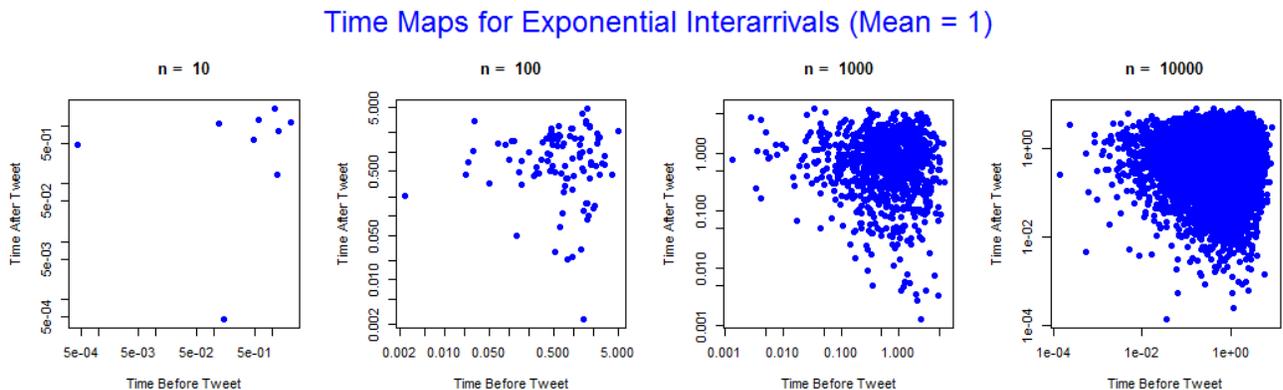

**Figure 3**: Time maps from simulated exponential interarrival streams with E(1) and varying sample sizes.

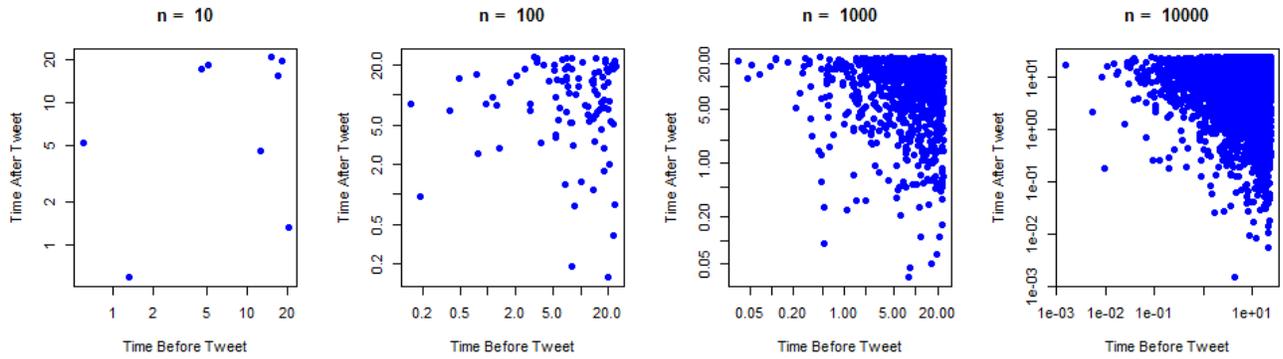

**Figure 4**: Time maps from simulated uniform interarrival streams with U(0,24) and varying sample sizes.

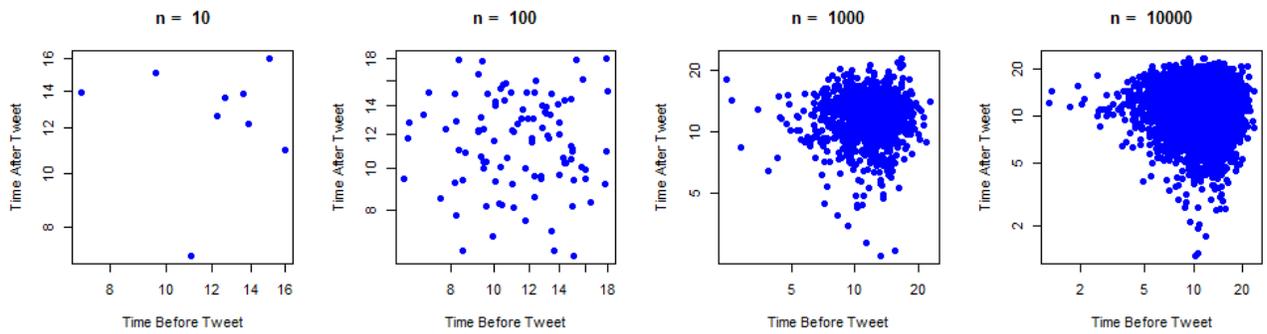

**Figure 5**: Time maps from simulated Gaussian interarrival streams with N(12,3) and varying sample sizes.

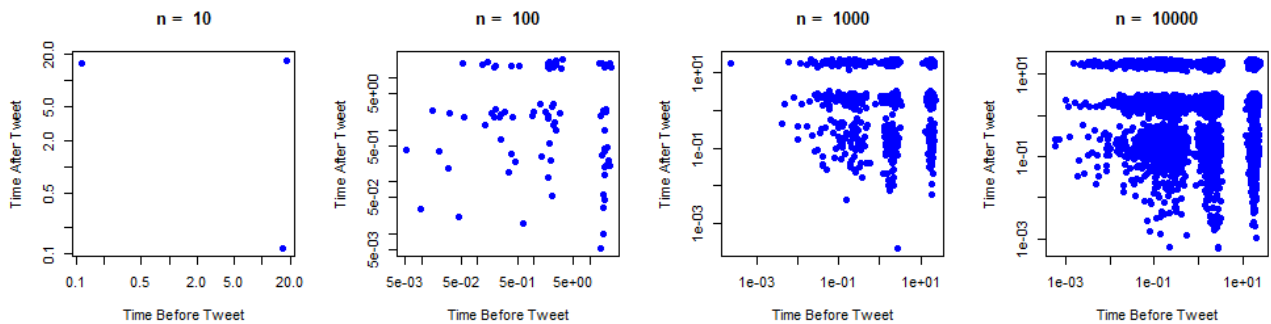

**Figure 6**: Time maps from a simulated mixture of Gaussian interarrival streams and varying sample sizes.

*Time Maps*

In addition to using the information generated in the simulations as a basis for understanding, Watson (2015) also associated regions of the time map with tweet behavior (Figure 7), partially based on his analysis of the @whitehouse Twitter account. Comparing the two, we might expect that human Twitter accounts would be more active in the center and upper right regions, consistent with exponential interarrival times, whereas maps for bots would favor the lower left. Furthermore, time maps for bots would tend to more broadly cover the entire space, reflecting the bot's consistency in posting on many timescales, unlike humans who are required to take breaks to sleep. We might also expect to see vertical or horizontal features in the interarrival maps for bots, which would correspond to excessive burstiness that is not likely to be observed from humans who are tweeting spontaneously.

|  | Last tweets of the day | Last tweets of the day |  |
|---|---|---|---|
| Bursts followed by lulls |  | Mundane events "Business as Usual" | First tweets of the day |
| Bursts followed by lulls | Major events |  | First tweets of the day |
| Extremely rapid bursts | Lulls followed by bursts | Lulls followed by bursts |  |

**Figure** 7: Watson (2015) associated regions of the (log scale) time map associated with specific activities.

To evaluate these hypotheses, we generated time maps from real tweet interarrival times (over the period December 6 - December 20, 2015) captured from the target users selected and listed in Table 1. These time maps are displayed in Figure 8 through Figure 12. Two variations of time maps are presented: 1) scatterplots (blue dots) and 2) heatmaps (red). The results are nearly identical, but the heatmaps are more effective at handling overplotting, and less consistent in axis scaling. These two approaches were used to see if additional information could be acquired via the visualizations, emphasizing a logarithmic scale to preserve Watson's intent that interarrival times on multiple timescales could be quickly compared using these maps. The scales on both blue scatterplots and red heatmaps are not consistent from plot to plot, so caution must be exercised when the plots are compared.

The plots were examined in groups of three or six: 1) known real humans, in Figure 8, 2) known real teenage girls compared to a bot who acts like a teenage girl in Figure 9, 3) known bots or humans who use tweet scheduling software in Figure 10, 4) Presidential Candidates in Figure 11, 5) users with a known strong social media strategy in Figure 12, and 6) a resampling of

@realdonaldtrump's 1787 tweets with n=114 for direct comparison to @hillaryclinton and @berniesanders with sample size removed as a contributing factor in Figure 13.

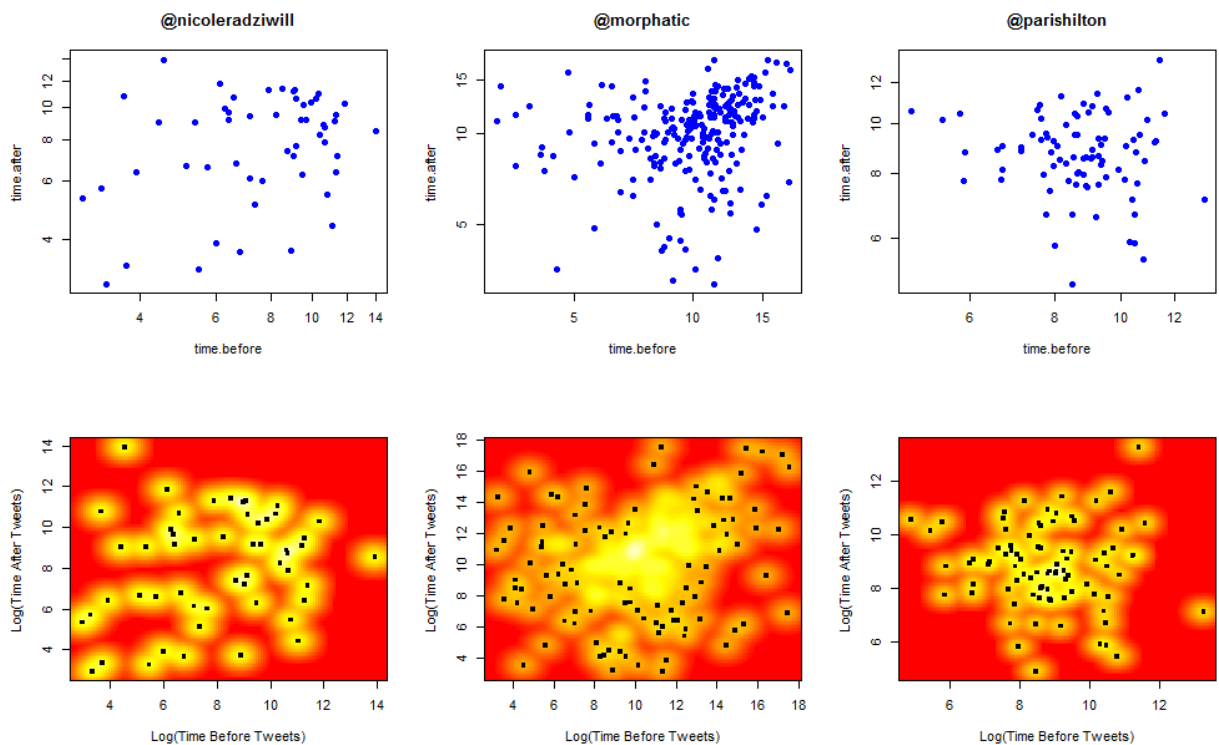

**Figure 8**: Time maps from humans who tweet spontaneously.

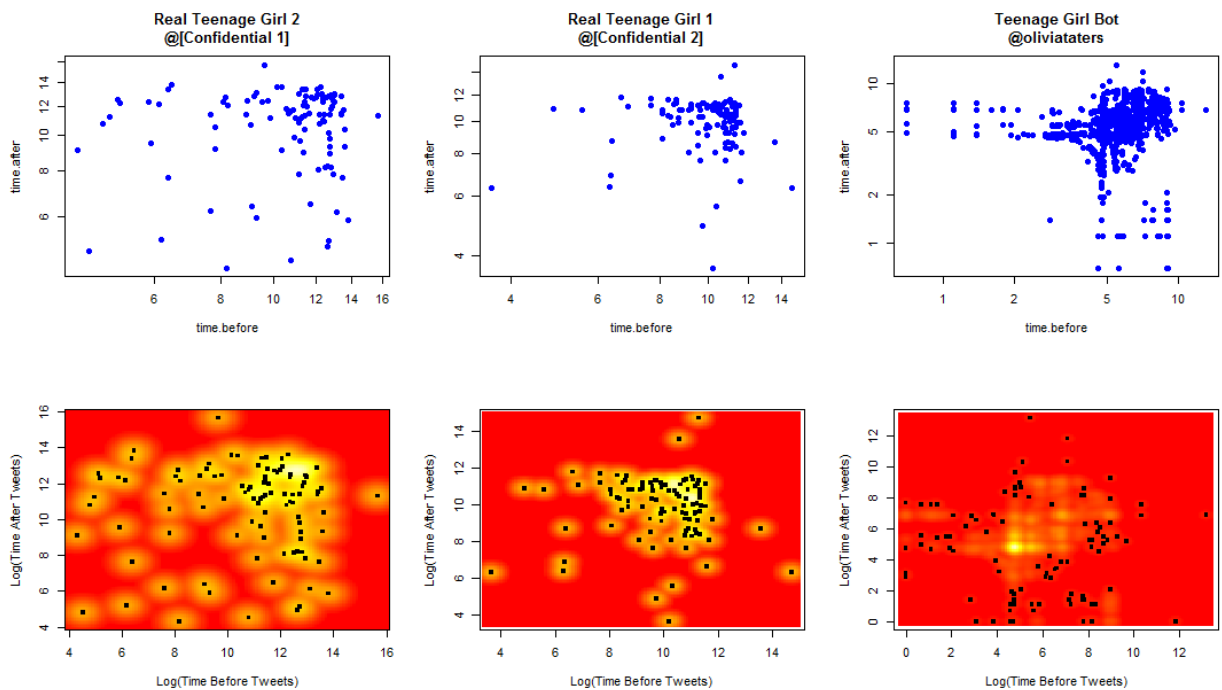

**Figure 9**: Time maps from human teenage girls compared to a simulated Teenage Girl Bot (@oliviataters).

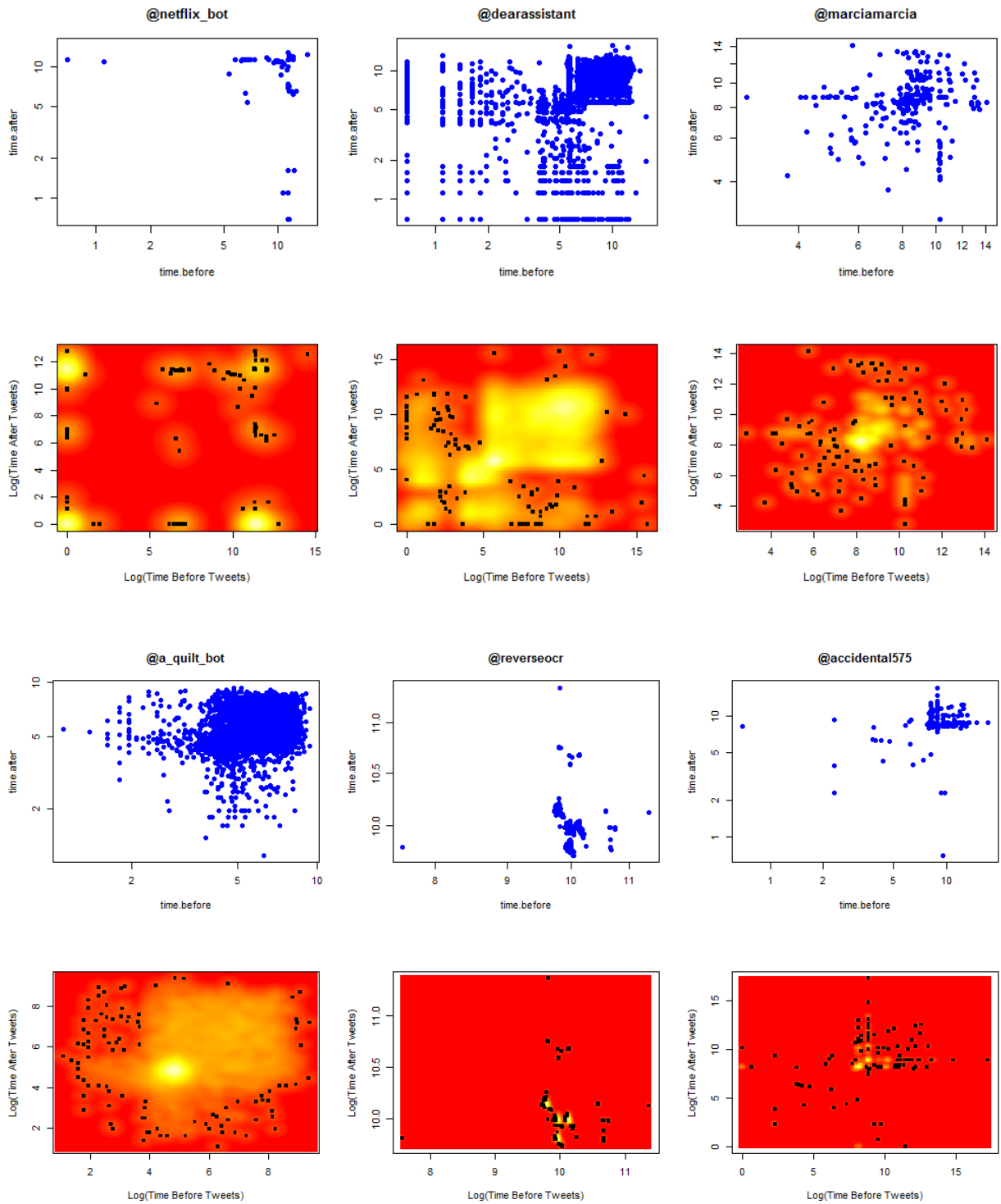

**Figure 10**: Time maps from bots and one human who schedules tweets (@marciamarcia).

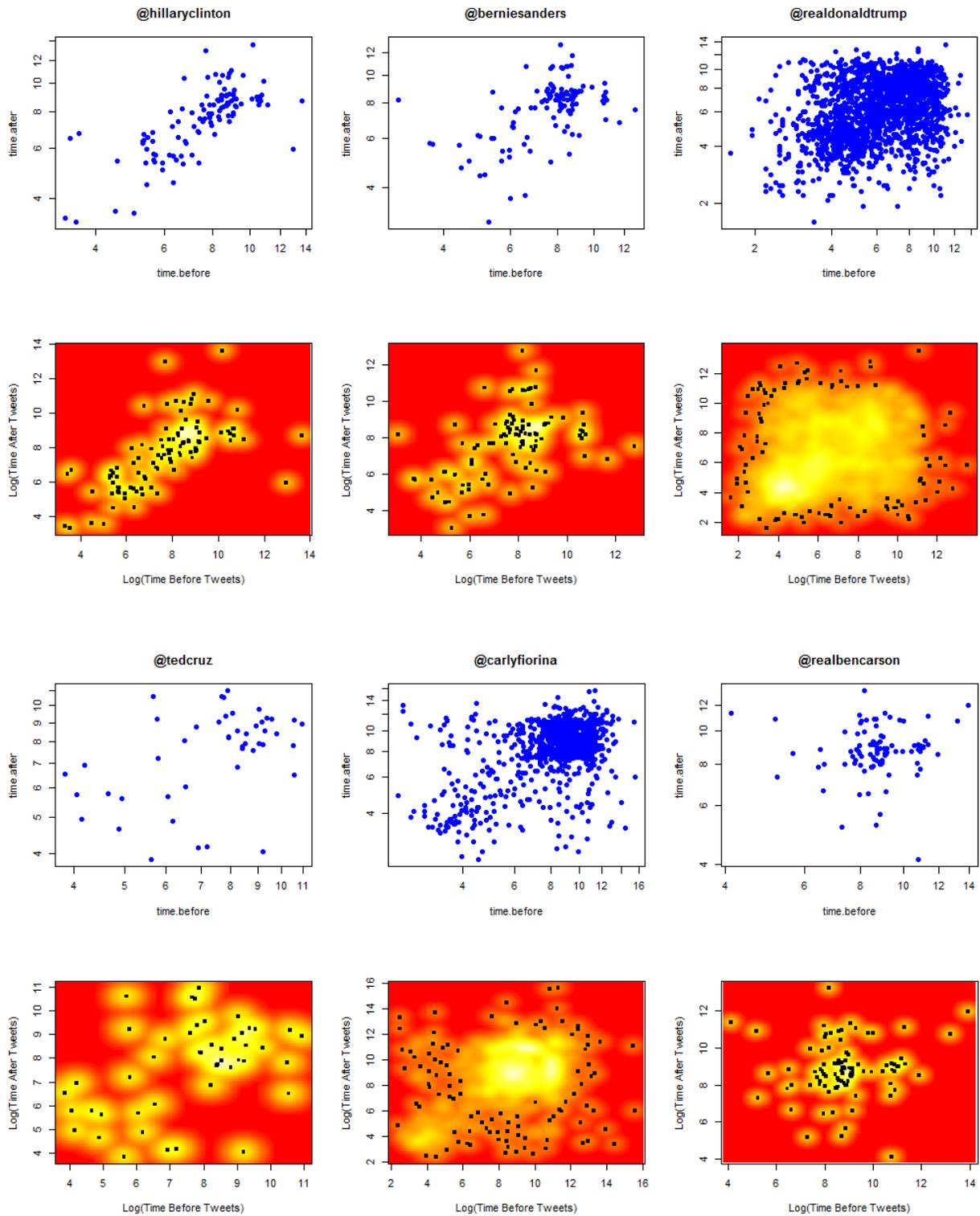

**Figure 11**: Time maps from a subset of the 2016 U.S. Presidential Candidates.

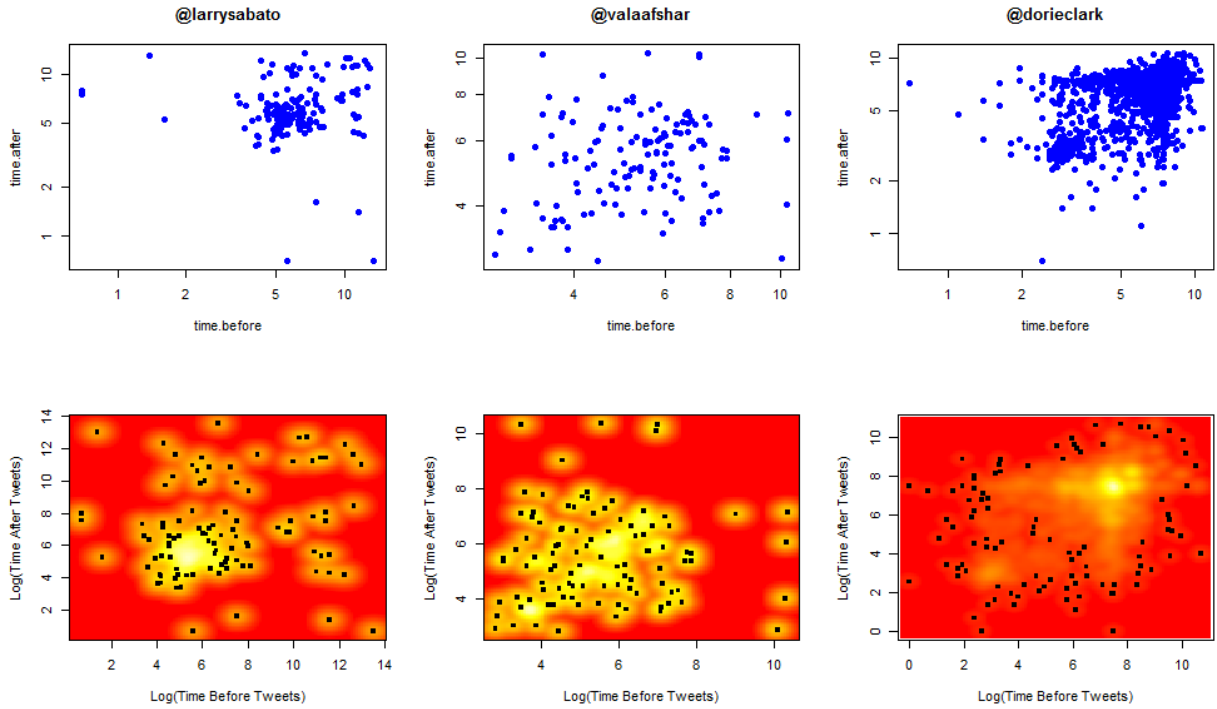

**Figure 12**: Time maps from target users with strong social media presence and strategy.

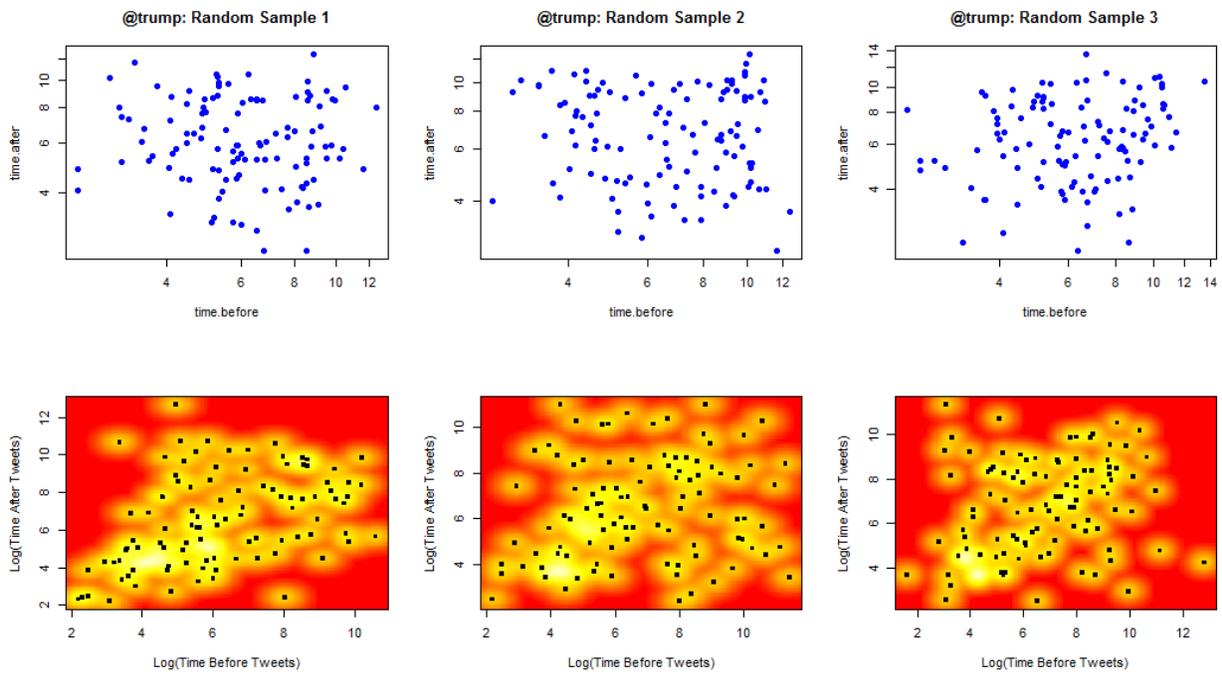

**Figure 13**: Resampling of @realdonaldtrump's 1787 tweets to n=114, for comparison with @hillaryclinton and @berniesanders.

*Discussion*

Based on examining the time maps for patterns (in particular, the simulation results in Figure 3 through Figure 6, and the patterns identified in Figure 7) we classified each of the target users based on whether we thought the account featured a human (spontaneously tweeting) or a bot (autonomously issuing tweets).

| Group | Twitter ID | Sample Size (n) | Observations | Possible DGP | Bot or Human? |
|---|---|---|---|---|---|
| **1: Humans who tweet spontaneously** | @nicoleradziwill | 54 | Lots of variability; no horizontal or vertical features, no clusters | Uniform | Human |
| | @morphatic | 233 | Lots of variability; no horizontal or vertical features, tweet clusters in both "mundane events" and "major events" | Uniform | Human |
| | @[Confidential 1] | 112 | Lots of variability; no horizontal or vertical features, tweets cluster in "mundane events" | Exponential | Human |
| | @[Confidential 2] | 118 | Lots of variability; no horizontal or vertical features, tweets cluster in "mundane events" | Exponential | Human |
| | @larrysabato | 149 | Lots of variability; no horizontal or vertical features, tweets primarily "major events" | Exponential | Human |
| **2: Humans with a social media presence** | @marciamarcia | 503 | Lots of variability with some horizontal and vertical features, no rapid bursts; cluster in "mundane events" | Uniform or Exponential | Human Using a Bot to Automate Tweets |
| | @valaafshar | 139 | Lots of variability with no horizontal or vertical features, but emphasis in lower left (rapid bursts) | Gaussian | Human Using a Bot to Automate Tweets |
| | @dorieclark | 3039 | Some variability; no horizontal or vertical features, clusters in "major events" as well as "mundane events" suggest frequent, spontaneous tweets | Exponential | Human |
| | @parishilton | 87 | Lots of variability; no horizontal or vertical features, cluster in "major events" is consistent with advertising events | Gaussian | Human |
| **3: Automated bots** | @dearassistant | 3160 | Many horizontal and vertical features; tweets primarily "mundane events" with many bursts and lulls | Mixture | Bot |
| | @oliviataters | 3200 | Many horizontal and vertical features; tweets primarily | Mixture | Bot |

| | | | "mundane events". Pattern is identical to real teenage girls in this sample. | | |
| --- | --- | --- | --- | --- | --- |
| | @a_quilt_bot | 3200 | Some horizontal and vertical features, no rapid bursts; cluster in "mundane events" | Uniform or Exponential | Bot |
| | @reverseocr | 1996 | Characterized by long lulls, then bursts; some vertical features; very limited variability | Gaussian (with small dispersion) | Bot with Unique Behavior |
| | @accidental575 | 818 | Limited variability, many horizontal and vertical features, does not tweet frequently | Gaussian (with small dispersion) | Bot with Unique Behavior |
| | @netflix_bot | 3200 | Limited variability, lots of horizontal and vertical features; infrequent, scheduled tweets | Uniform | Bot |
| **4: 2016 U.S. Presidential Candidates** | @berniesanders | 114 | Lots of variability; no horizontal or vertical features, tweet clusters in both "mundane events" and "major events" | Gaussian | Human |
| | @hillaryclinton | 114 | Lots of variability; no horizontal or vertical features, tweet clusters in both "mundane events" and "major events" | Gaussian | Human |
| | @realdonaldtrump | 1787 | Lots of variability, no detectable horizontal or vertical features because tweet frequency extremely high. Resampling to n=114 suggests that the pattern is markedly different than for competitors @hillaryclinton and @berniesanders. | Gaussian | Bot |
| | @realbencarson | 97 | Limited variability, no horizontal and vertical features, cluster in "mundane events" so possibly avoids tweeting about major events | Gaussian | Human |
| | @carlyfiorina | 1089 | Lots of variability; no horizontal or vertical features, tweets balanced with clusters in both "mundane events" and "major events" | Gaussian | Human |
| | @tedcruz | 48 | Lots of variability; no horizontal or vertical features, tweets cluster in "mundane events" | Exponential | Human |

**Table 2.** Patterns observed in the time maps for the target users..

Several themes were noticed while inspecting these time maps, given that a vertical feature indicates a lull followed by a burst, and a horizontal feature shows a burst followed by a lull:

- The time maps for every known bot featured clear horizontal and/or vertical features that correspond to long lulls and high-frequency bursts. Spontaneous human tweets did not demonstrate this pattern.
- The time maps for users who are known to be strong social media strategists were associated with a cluster in the mid lower left quadrant, indicating that they tweet more frequently than other humans, but not as frequently as bots who produce tweets in rapid bursts.
- Gaps along the top margin and the right margin appear to indicate several long lulls ended by low-frequency bursts, and appeared to be unique among the human Twitter users in this sample. We suspect that these are the gaps that result from a sleep pattern.
- The most bursty time map was associated with "Data Assistant" (@dataassistant), an Twitter account that receives questions and attempts to answer them using Wolfram Alpha. The time map suggests that once the computation is performed, this account can tweet many responses back to the querent in rapid succession, rather than just one.
- Several accounts were associated with broad and diffuse patterns on the heat maps, within which no distinguishable horizontal or vertical features were apparent (@realdonaldtrump, @dearassistant, @a_quilt_bot, @carlyfiorina, @dorieclark). It was not clear what these accounts had in common that might yield this pattern.

Five bots were selected for this exploratory study to represent each of the five characteristics of intelligent agents originally outlined by Franklin & Graesser (1997). Twitter accounts which could be classified as intelligent systems revealed time maps that were both human-like and very non-human:

- Time map analysis showed that the most bot-like behavior is exhibited by the **adaptive** system (@dearassistant) that learns from its questions and can respond to the querent many times in succession. The **proactive** (@oliviataters) and **interactive** (@a_quilt_bot) accounts were associated with time maps that appeared the most human, which is not surprising since the goal of @oliviataters is to *act* human, and @a_quilt_bot cannot respond unless it is initiated by a human request. The **autonomous** (@reverseocr) and **reactive** (@accidental575) accounts exhibited behavior that was not like humans or other bots; this behavior could be the subject of future study.
- Time maps for "Reverse OCR" (@reverseocr) and "Accidental Haiku" (@accidental575) bots showed extremely limited variability in tweet interarrival times. Reverse OCR is a program that generates random line segments until its character recognition software detects a word, and then it tweets the line segments and the word. Accidental Haiku is a program that searches for tweets that are written using the haiku literary structure (5 syllables, 7 syllables, 5 syllables) and then retweets them. These accounts are engaged in a continuous search process that continues until a solution is found, which is distinctly different that a spontaneous tweet process or a scheduled tweet process.

The pattern observed with U.S. Presidential candidates' time maps included the following:

- The time maps for Bernie Sanders (@berniesanders) and Hillary Clinton (@hillaryclinton) are strikingly similar, and (like @carlyfiorina) indicate attention is being given to tweeting on a regular basis, frequently but not *too* frequently (like @valaafshar), and balancing coverage between mundane events and major or emerging events.
- Time maps from resampling tweets from Donald Trump's (@realdonaldtrump) original sample of 1787 to sample sizes of n=114 (matching the sample sizes for competitors @hillaryclinton and @berniesanders) show a consistent pattern: large variability, with a cluster in the lower left quadrant, corresponding to tweets about major events. The pattern is distinctly different than the pattern that was observed in the time maps from @hillaryclinton and @berniesanders.
- Time maps from U.S. Presidential Candidates Hillary Clinton (@hillaryclinton) and Bernie Sanders (@berniesanders) were nearly identical, and exhibited markedly different patterns than any of the other time maps that were considered.
- Paris Hilton (@parishilton) appears to be a human, and tweets most frequently about major events that are either upcoming or in progress, according to the time map.
- Larry Sabato (@larrysabato) appears to be a human, despite an admirable and aggressive practice of extremely frequent tweets focusing on both mundane and major events.
- Donald Trump (@realdonaldtrump) appears to be a bot. We will refrain from speculating about what this might mean, and will instead leave further investigation up to the reader.

Although qualitative analysis led to some curious results, a more comprehensive and rigorous quantitative analysis should be undertaken before these maps are used for any practical purpose.

**Conclusions and Future Work**

In this exploratory study, Watson's (2015) time maps were produced for 21 Twitter accounts, consisting of humans who tweet spontaneously, humans who use tweet using scheduling software, humans with strong social media presences, bots which demonstrate each of Franklin & Graesser's (1997) characteristics of intelligent agents, and 2016 U.S. Presidential Candidates from both the Democratic and Republican parties. Results indicate that it is usually easy to distinguish a bot from a human, so long as the human is tweeting spontaneously (and not using software to schedule tweets in advance). Some bots exhibited behavior that was not like other humans or other bots.

The method explored here is intriguing, but in its current state it is neither generalizable nor robust due to several limitations. Most significantly, time maps need to be examined for a much larger sample of users, and metrics should be expressed to quantitatively judge the differences (if any) between the profiles. In addition, this paper examined interarrival times only, and did not explore spatio-temporal aspects of tweet generation or even diurnal variations in tweet activity. Furthermore, a theoretical basis for the distinction between the patterns should be articulated. Supervised machine learning methods should be applied to determine the most accurate approach to automatically distinguish between humans, bots, and hybrids.

As a result, there are several research questions that naturally follow from this exploratory study, including:

- Is there a difference between autocorrelated interarrival streams (that is, whether future interarrivals depend on previous interarrival times) vs. those that are independent and identically distributed (i.i.d.)?
- What data generating process (DGP) informs temporal patterns in tweet activity? Is the DGP different between humans, bots, and hybrids?
- Is the arrival process that governs spontaneous tweets by humans usually exponential?
- Is there any relationship between the time of day tweets are generated, or the geographical location of the account holder, and the pattern on the time map?
- Can an approach based on time maps be used to determine what purpose someone is using their Twitter account for, and if this purpose changes suddenly?

The answers to these questions have implications for social media intelligence, continued development of intelligent agents who broadcast information (either autonomously or in response to human requests), and even potentially security. Although more research is required to determine concise models or heuristics based using time maps that would reliably predict what type of user a Twitter account is associated with, early results do suggest the presence of discriminating patterns.

## Appendix A: R Code Used to Generate Plots

```
logtimemap <- function(x,n) {
   # takes a list of INTERARRIVAL times, shifts it into x and y, and plots
   time.before <- x[-1]
   time.after <- head(x[1:length(x)],-1)
   plot(time.before,time.after,col="blue",pch=16,log="xy",
      xlab="Time Before Tweet", ylab="Time After Tweet", main=paste("n = ",n))
}

timemap.plus <- function(x,s) {
   # same as timemap but PLUS custom titles and fixed axis limits
   # takes a list of interarrival times, shifts it into x and y, plots
   time.before <- x[-1]
   time.after <- head(x[1:length(x)],-1)
   plot(time.before,time.after,col="blue",pch=16,main=paste(s),log="xy")
}

# EXPONENTIAL
par(mfrow=c(1,4),oma=c(0,0,5,0))
logtimemap(rexp(10),10)
logtimemap(rexp(100),100)
logtimemap(rexp(1000),1000)
logtimemap(rexp(10000),10000)
mtext("Time Maps for Exponential Interarrivals (Mean = 1)", side=3, line=0, adj=0.5, cex=1.5, col="blue", outer=TRUE)

# UNIFORM
par(mfrow=c(1,4),oma=c(0,0,5,0))
logtimemap(runif(10,0,24),10)
logtimemap(runif(100,0,24),100)
logtimemap(runif(1000,0,24),1000)
logtimemap(runif(10000,0,24),10000)
mtext("Time Maps for Uniform Interarrivals (Min = 0, Max = 24)", side=3, line=0, adj=0.5, cex=1.5, col="blue", outer=TRUE)
```

```r
# NORMAL
par(mfrow=c(1,4),oma=c(0,0,5,0))
logtimemap(rnorm(10,mean=12,sd=3),10)
logtimemap(rnorm(100,mean=12,sd=3),100)
logtimemap(rnorm(1000,mean=12,sd=3),1000)
logtimemap(rnorm(10000,mean=12,sd=3),10000)
mtext("Time Maps for Gaussian Interarrivals (Mean = 12, SD = 3)", side=3, line=0,
adj=0.5, cex=1.5, col="blue", outer=TRUE)

# MIXED GAUSSIAN
mixture <- function(N) {
   components <- sample(1:3,prob=c(0.33,0.33,0.33),size=N,replace=TRUE)
   mus <- c(.12,2,18)
   sds <- sqrt(c(.03,.2,3))
   samples <- rnorm(n=N,mean=mus[components],sd=sds[components])
}

par(mfrow=c(1,4),oma=c(0,0,5,0))
logtimemap(mixture(10),10)
logtimemap(mixture(100),100)
logtimemap(mixture(1000),1000)
logtimemap(mixture(10000),10000)
mtext("Time Maps for Mixed Gaussian Interarrivals", side=3, line=0, adj=0.5, cex=1.5,
col="blue", outer=TRUE)

combo3.tm <- function(otdf1,xy1,who1,otdf2,xy2,who2,otdf3,xy3,who3) {
   # Produce 6-panel plots with 3 users, top = a LOG-LOG plot of logged data
   par(mfrow=c(2,3))
   timemap.plus(log(otdf1$secondsSincePrevious),who1)
   timemap.plus(log(otdf2$secondsSincePrevious),who2)
   timemap.plus(log(otdf3$secondsSincePrevious),who3)
   Lab.palette <- colorRampPalette(c("red", heat.colors(10)), space = "Lab")
   smoothScatter(log(xy1), colramp = Lab.palette, cex=6,
      xlab="Log(Time Before Tweets)", ylab="Log(Time After Tweets)")
   smoothScatter(log(xy2), colramp = Lab.palette, cex=6,
      xlab="Log(Time Before Tweets)", ylab="Log(Time After Tweets)")
   smoothScatter(log(xy3), colramp = Lab.palette, cex=6,
      xlab="Log(Time Before Tweets)", ylab="Log(Time After Tweets)")
}
```

## Appendix B: R Code Used to Obtain Tweet Interarrival Times

```r
function loadTweets(accountname) {
   install.packages('twitteR')
   library(twitteR)

   # get authorization
   setup_twitter_oauth(consumer_key = "Insert your key from http://apps.twitter.com",
   consumer_secret = "Insert your key from http://apps.twitter.com")

   # download the most recent 0-3200 tweets from @accountname
```

```r
    ot_tweets = userTimeline(accountname, n = 3200)

    # create a function to add each tweet to data frame
    addTweet <- function(index, tweets, df) {
        if (index < 3199) {
           secondsSincePrevious = as.integer(difftime(tweets[[index]]$getCreated(),
           tweets[[index + 1]]$getCreated(), units = "secs"))
        } else {
           secondsSincePrevious = 0
        }
        if (index > 1) {
           secondsUntilNext = as.integer(difftime(tweets[[index - 1]]$getCreated(),
           tweets[[index]]$getCreated(), units = "secs"))
        } else {
           secondsUntilNext = 0
        }
            newrow = list(created=tweets[[index]]$getCreated(),
secondsSincePrevious=secondsSincePrevious,secondsUntilNext=secondsUntilNext)
        df = rbind(df, newrow)
        return(df)
      }

# create empty data.frame with columns to be added
otdf <- data.frame(created=as.Date(character()),
secondsSincePrevious=integer(),secondsUntilNext=integer())

# calculate seconds since previous tweet, and seconds until next tweet
for(i in 1:3200) { otdf = addTweet(i,ot_tweets,otdf)}
return(otdf)
}
```